# Stability studies of nanosecond light sources based on blue ultra bright LEDs.

Lubsandorzhiev[*] B.K., Vyatchin Y.E.

*Institute for Nuclear Research of RAS*

[*]Corresponding author:

Tel: +7(095)1353161; fax: +7(095)1352268;

e-mail: lubsand@pcbai10.inr.ruhep.ru

Address: 117312 Moscow Russia, pr-t 60-letiya Oktyabrya 7A

Institute for Nuclear Research of RAS

## Abstract

*We present the results of stability studies of nanosecond light sources based on single quantum well (SQW) InGaN/GaN ultra bright blue LEDs. It is shown that the light yield of such light sources and their timing characteristics don't deteriorate even after $10^{10}$ total pulses. The longterm stability of the sources light yield is better than 1%.*

The advent of ultra bright blue LEDs based on InGaN/GaN [1,2] structures at the beginning of 1990s opened new era in the development of nanosecond light sources for use in different fields of experimental physics: time and amplitude calibrations of Cherenkov and scintillator detectors, fluorescent measurements, studies of fast processes kinetics [3-7] etc.

To drive such LEDs two types of drivers are widely used. One of them is so called "Kapustinsky's driver" [8] which uses a fast discharge of a small capacitor via a complementary pair of fast transistors. The highest level of light yield and at the same time the shortest light pulses are reached with another driver type which exploits avalanche transistors to have very short, nanosecond width current pulses flowing through LED with amplitudes of up to 3A [7]. Drivers based on avalanche transistors are currently used in a number of astroparticle physics experiments such as the lake Baikal neutrino experiment [9], Cherenkov EAS detector TUNKA [10], MAGIC Telescope [11] etc.

Currently a plethora of ultra bright blue LEDs are available in the market. We have tested LEDs from 20 manufacturers: NICHIA, KINGBRIGHT, Agilent Technologies, LIGITEK, Bright LED, LUXPIA, COTCO and YolDal among them. For stability measurements we have chosen LEDs produced by NICHIA CHEMICAL Ltd and KINGBRIGHT Corporation companies – NSPB500S (and/or NSPB300S) and L7113NBC (and/or L-53NBC) correspondingly. They are ultra bright SQW InGaN/GaN LEDs with ~1÷2 cd light brightness [12,13]. The main reason to choose these two types of LEDs was that they were the fastest among other ultra bright blue LEDs tested by us.

The longterm stability and the temperature dependence of the light yield of nanosecond light sources based on the blue ultra bright LEDs and an avalanche transistor driver were studied with the set-up presented in Fig.1. The light source under studies comprises LED and an electrical driver. The driver scheme is shown in Fig.2. The avalanche transistors are FMMT415 manufactured by ZETEX. The discharge capacitor C3 was adjusted in the range of 3-10 pF to have current pulses flowing through the LED with amplitudes of 0.6-2.2 A.

A light source under studies is put into a thermostatic chamber with temperature stabilization of $\pm 0.1^0 C$. The thermostate temperature is adjustable with $1-3^0 C$ step. To illuminate a photomultiplier (PMT) photocathode by light pulses from the light source arranged in the thermostate we have used a short optical cable (*F* in Fig.1). A PMT used in the set-up was ET9116B developed by Electron Tubes Ltd for MAGIC Telescope project [14]. ET9116B is a fast six stage PMT with a hemispherical photocathode and a few hundred picoseconds jitter. The PMT's anode pulses are amplified by a fast transimpedance preamplifier and LeCroy 612 amplifier. We have used LeCroy 2249A unit as ADC to measure charges of the PMT anode pulses. BVT-12A [15] type of TDC with 75 ps step and 5 µs range and constant fraction discriminators were used to measure light pulses profiles. A set of neutral density filters was used to have ~100 photoelectrons illumination of the PMT photocathode for charge measurements and a single photoelectron level of illumination to measure the LEDs light emission kinetics.

To control stability of the whole measurement system a calibration source [16] ($R$ in Fig.1) made of a thin layer of YAlO$_3$:Ce scintillator covered with Am$^{241}$ isotope was optically coupled to the PMT window. This calibration light source provides pulses with amplitude of ~1000 p.e. on the PMT photocathode with ~100 s$^{-1}$ counting rate. The long term stability measurements were carried out at room temperature. The level of fluctuations of the whole spectrometric channel gain including the PMT gain is less than ~1 %, see Fig.3a where we show for a convenience reason the dependence of the gain versus the total number of the light source pulses. The width and amplitude of current pulses flowing through LED were controlled as well and it's noteworthy to mention here they were stable throughout the measurement.

We have measured the longterm stability at three amplitudes of current pulses running through LED - 0.6A, 1.2A and 2.2A. Besides the measurements were carried out at four repetition rates of the light source output pulses: 100Hz, 1kHz, 10kHz and 100 kHz for each value of current amplitude. The results were practically the same for all currents and repetition rates. The results of the longterm stability measurements of the NICHIA and KINGBRIGHT LEDs are presented in Fig.3b and 3c respectively. In this paper we present the data gained under the harshest conditions – 2.2A current pulses and 100 kHz repetition rate. One can see the LEDs and driver parameters are very stable. The light yields of the sources did not show any deterioration even after $10^{10}$ total output pulses. Furthermore the fluctuations of the measurement system gain contribute substantially to the total fluctuations. In case of the KINGBRIGHT LEDs the fluctuations are a bit higher in comparison with the NICHIA LEDs but still less than 1% (compare Fig.3a with 3b and 3c).

The temperature dependencies of the light sources based on the NICHIA and KINGBRIGHT LEDs are shown in Fig.4a and 4b respectively. It is interesting that the two types of LEDs demonstrate different dependencies on temperature in the range of $-3^0C \div +45^0C$. The NICHIA LEDs light yield increases by ~7% with temperature rise from $-3^0C$ up to $+45^0C$. The temperature dependence of the LED light yield is approximated well by the following expression:

$$\mathbf{Y = Y_0(1.12 - 0.15e^{\alpha T})}, \qquad (1)$$

where $\alpha = -0.015\ ^0C^{-1}$ – temperature coefficient, T – temperature in $^0C$.

In contrary with the NICHIA LEDs the KINGBRIGHT LEDs have inverse temperature behavior in the same temperature range. Their light yield decreases by ~9% when temperature increases from $-3^0C$ up to $+45^0C$ and the light yield dependence on temperature is described by the following formulae:

$$\mathbf{Y = Y_0\ (0.94 + 0.105e^{\alpha T})}, \qquad (2)$$

where $\alpha = -0.026 \,^0C^{-1}$ – temperature coefficient, **T** – temperature in $^0C$.

We have controlled the light pulses temporal profiles during the longterm and temperature stability measurements. The dependencies of the light pulses width (FWHM) on the total number of the light source output pulses and temperature are shown in Fig.5a and 5b respectively. One can see that the light pulses width didn't notably alter throughout the measurements.

To conclude nanosecond light sources based on SQW InGaN/GaN ultra bright blue LEDs NICHIA NSPB500S (NSPB300S) and KINGBRIGHT L7113NBC (L-53NBC) demonstrates very high longterm stability. Their light yields and timing characteristics remain practically the same after $10^{10}$ pulses even under very harsh conditions: current pulses amplitude up to 2.2A and 100 kHz repetition rates. Quite unexpectedly the two types of LEDs show different behaviors of their light yields temperature dependencies in the temperature range $-3^0C$-+$45^0C$ but their timing characteristics are invariable in this temperature range. Altogether we have studied 15 LEDs of each type in our measurements. Characteristics of all studied LEDs had very good repetitiveness. Here we have presented typical results.

Unfortunately it should be noted here that for the NICHIA ultra bright blue LEDs, even for the same NSPB300 or NSPB500 type of LEDs, only the "old" ones produced in the middle of 1990s show very fast light emission kinetics. One can see in Fig.6 the NICHIA NSPB500S LEDs light pulses profiles measured with the set-up described above. It is clearly seen that there are two batches of LEDs. The first "fast" group consists of LEDs bought in 1995-96 (we call them "old" LEDs) and the second one corresponds to LEDs bought in 2002-03 ("new" LEDs). As to the KINGBRIGHT ultra bright blue LEDs family only LEDs with NBC mark extension – L7113NBC and L-53NBC, which are unfortunately obsolete and presently not available, demonstrate fast light emission kinetics similar to kinetics of the NICHIA "old" LEDs. New LEDs presently available from KINGBRIGHT (e.g. L7113PBC and L-53PBC) are conspicuously slower as it can be seen in Fig.7. It is beyond the scope of the present work to discuss the causes of such drastic changes in the LEDs light emission kinetics. It will be discussed more thoroughly elsewhere [17].

We are indebted very much to Dr. E.Lorenz for initiating the present work and Dr.V.Ch.Lubsandorzhieva for careful reading of this article and many invaluable remarks. The work has been done due to support of Russian Foundation of Basic Research grant #02-02-17365.

Figures.

Fig.1. Experimental set-up for LEDs longterm stablity studies.

*F* – optical cable; *R* – calibration light source (Am$^{241}$ on YAlO$_3$ scintillator); *Attenuators* – set of neutral density optical filters; *PMT* – ET9116B; *Preamp* – transimpedace preamplifier; *Amp* – amplifier LeCroy 612AM; *Delay* – delay line; *CFD$_1$* and *CFD$_2$* – constant fraction discriminators; *Generator* – pulse generator Stanford DG535; *&* - coincidence unit LeCroy 466.

Fig.2. Electrical scheme of LED driver.

Fig.3. Dependencies of the system gain (a), NICHIA NSPB500S (b) and KINGBRIGHT L7113NBC (c) light yields on the total number of the light source output pulses.

Fig.4. Dependence of the LEDs light yield on temperature: a) – for NICHIA NSPB500S; b) – for KINGBRIGHT L7113NBC.

Fig.5. Dependence of NICHIA NSPB500S light pulses width on the total number of the light source pulses (a) and temperature (b).

Fig.6. Light pulses temporal profiles for "old" and "new" NICHIA ultra bright blue LEDs, curves 1 and 2 respectively.

Fig.7. Light pulses temporal profiles for "old" (L7113NBC) and "new" (L7113PBC) Kingbright ultra bright blue LEDs, curves 1 and 2 respectively.

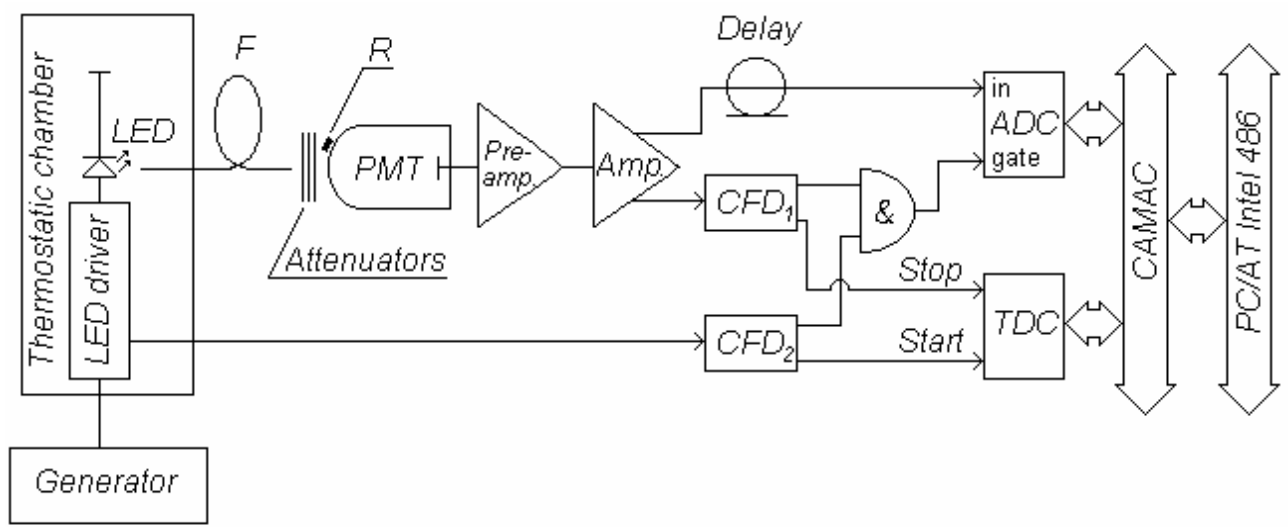

Figure 1.

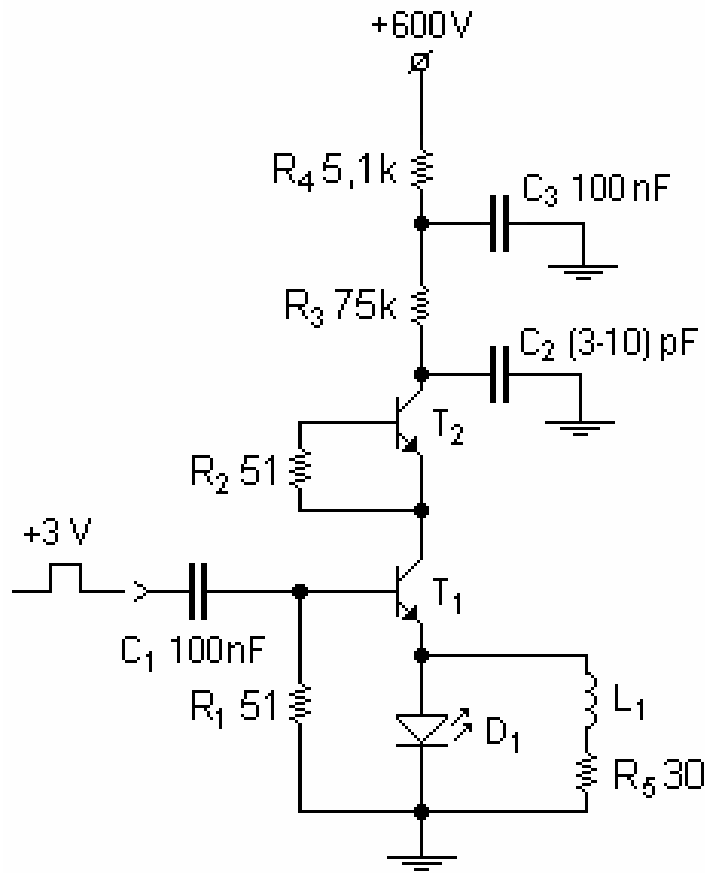

Figure 2.

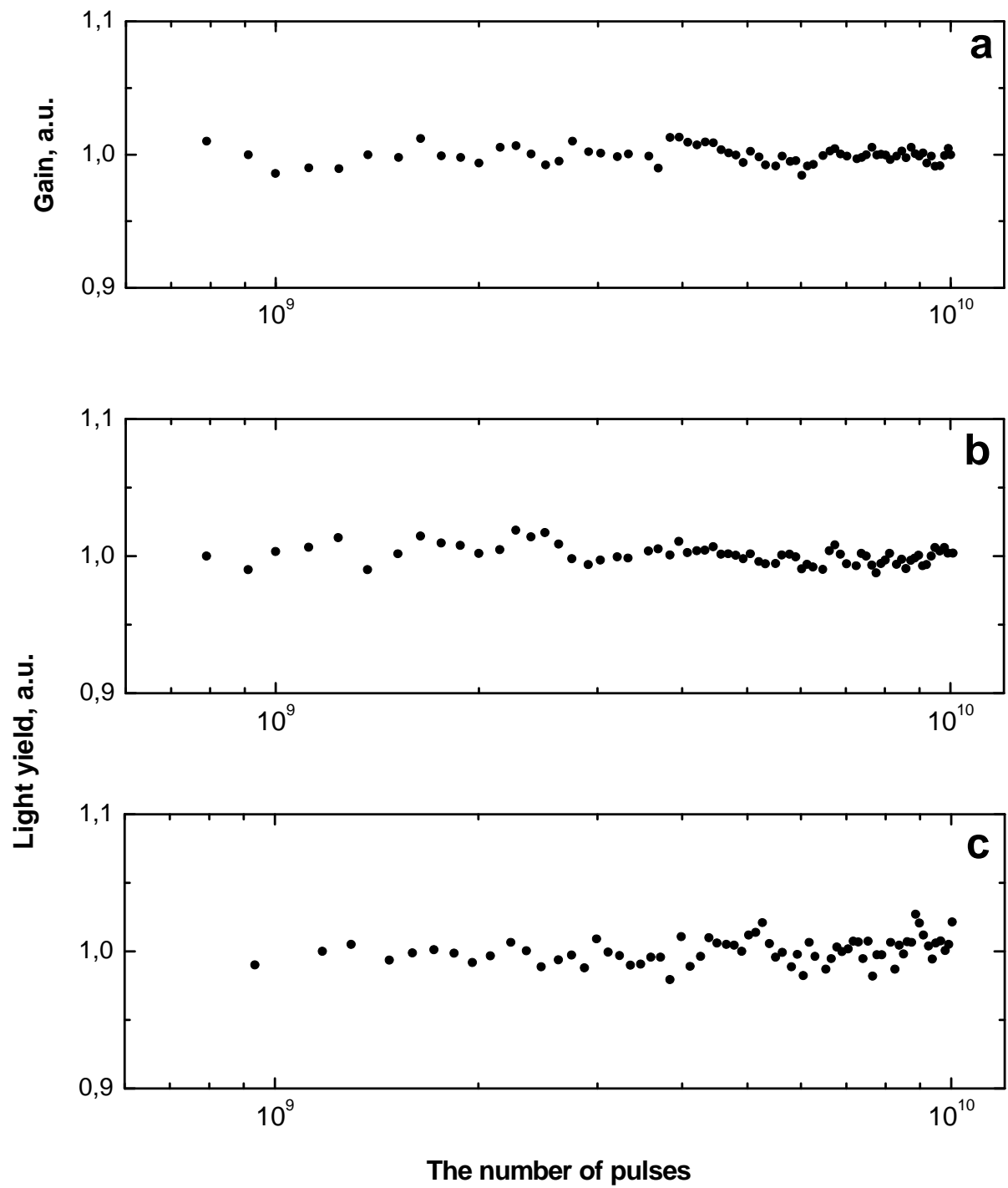

Figure 3

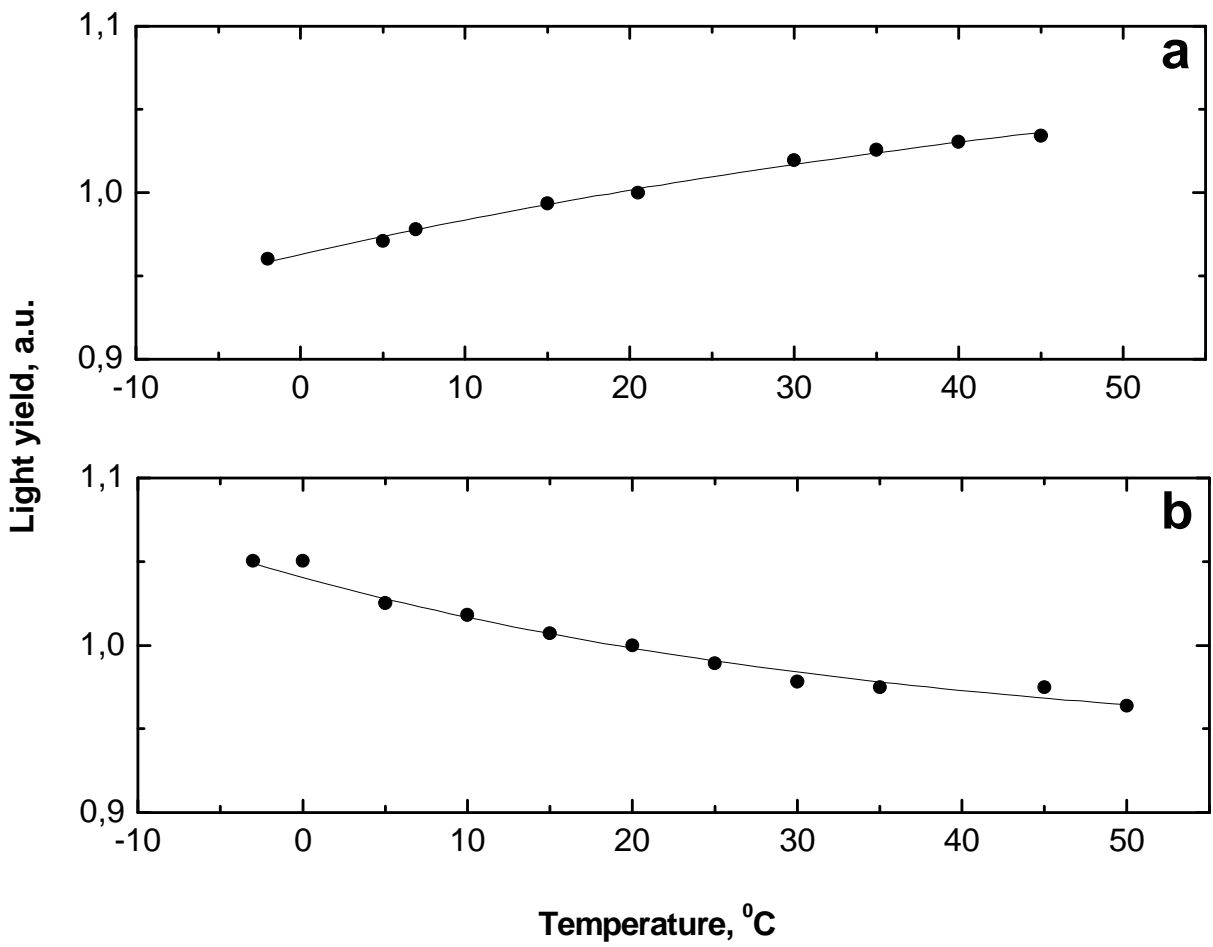

Figure 4.

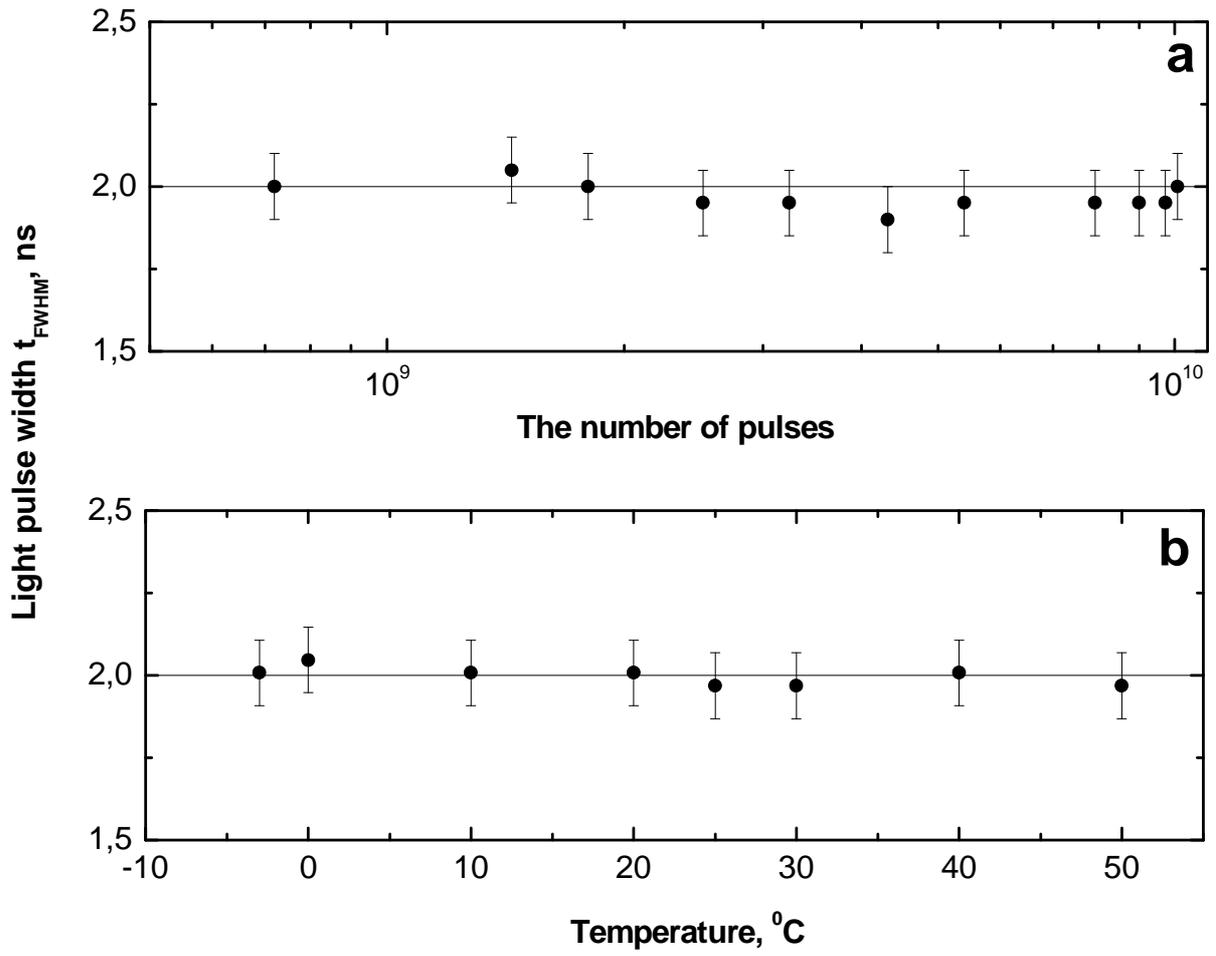

Figure 5.

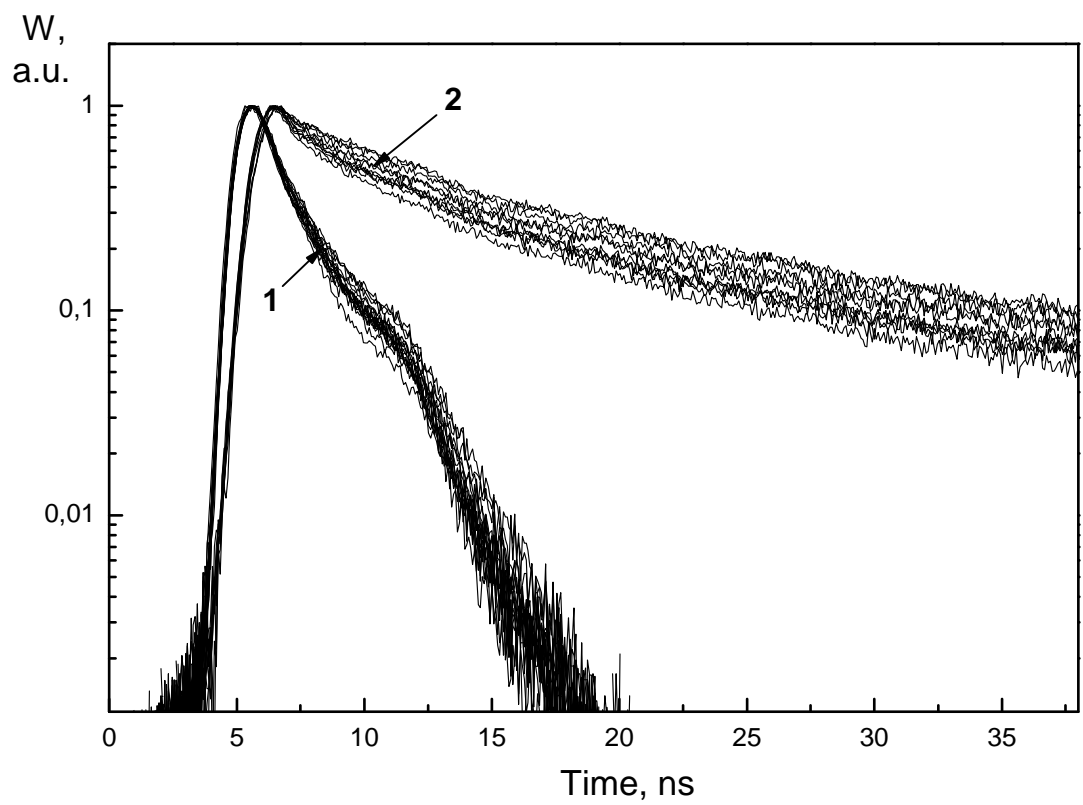

Figure 6.

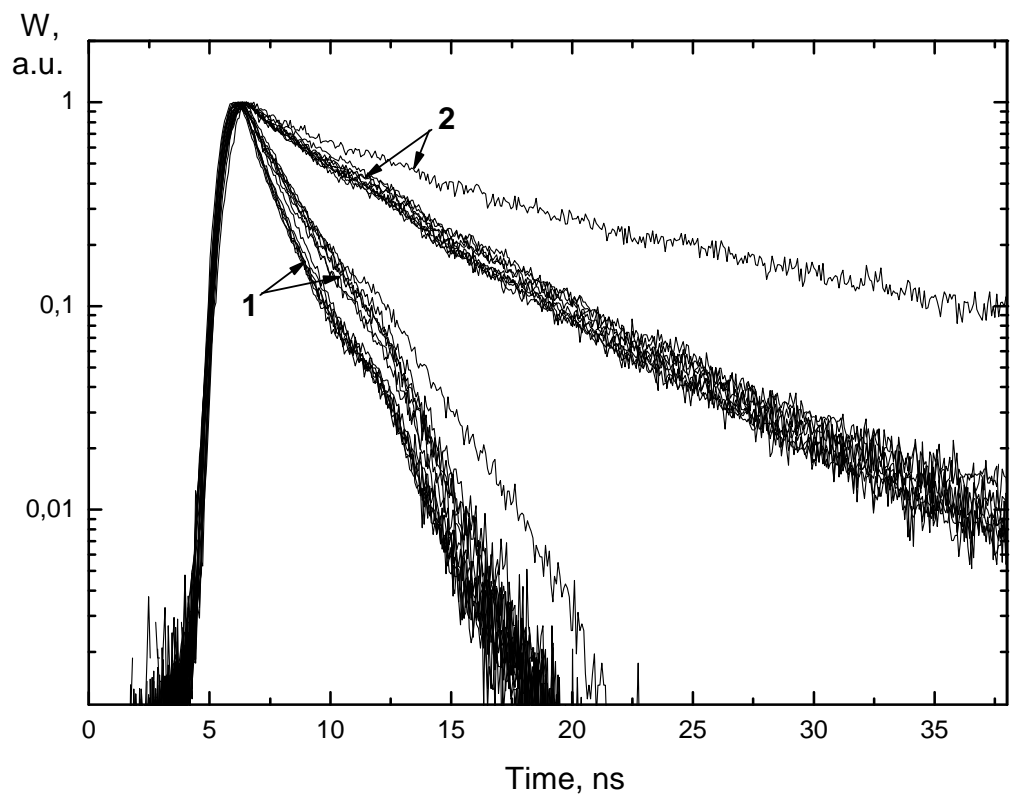

Figure 7.